\documentclass[conference]{IEEEtran}
\IEEEoverridecommandlockouts
\usepackage{cite}
\usepackage{amsmath,amssymb,amsfonts}
\usepackage{algorithmic}
\usepackage{graphicx}
\usepackage{textcomp}
\usepackage{balance}
\usepackage{xcolor}
\def\BibTeX{{\rm B\kern-.05em{\sc i\kern-.025em b}\kern-.08em
    T\kern-.1667em\lower.7ex\hbox{E}\kern-.125emX}}
\begin{document}

\title{Quality Assurance for LLM-RAG Systems: Empirical Insights from Tourism Application Testing

}

\author{\IEEEauthorblockN{Bestoun S. Ahmed}
\IEEEauthorblockA{\textit{dept. Mathematics and Computer Science} \\
\textit{Karlstad University}\\
Karlstad, Sweden \\
bestoun@kau.se}
\and
\IEEEauthorblockN{Ludwig Otto Baader}
\IEEEauthorblockA{\textit{dept. Mathematics, Informatics and Statistics} \\
\textit{Ludwig Maximilian University Munich}\\
Munich, Germany \\
ludwig.baader@googlemail.com}
\and
\IEEEauthorblockN{Firas Bayram}
\IEEEauthorblockA{\textit{dept. Mathematics and Computer Science} \\
\textit{Karlstad University}\\
Karlstad, Sweden \\
firas.bayram@kau.se}

\and
\IEEEauthorblockN{Siri Jagstedt}
\IEEEauthorblockA{\textit{CTF, Service Research Center
} \\
\textit{Karlstad University}\\
Karlstad, Sweden \\
siri.jagstedt@kau.se}

\and
\IEEEauthorblockN{Peter Magnusson}
\IEEEauthorblockA{\textit{CTF, Service Research Center
} \\
\textit{Karlstad University}\\
Karlstad, Sweden \\
peter.magnusson@kau.se}

}

\maketitle

\begin{abstract}
This paper presents a comprehensive framework for testing and evaluating quality characteristics of Large Language Model (LLM) systems enhanced with Retrieval-Augmented Generation (RAG) in tourism applications. Through systematic empirical evaluation of three different LLM variants across multiple parameter configurations, we demonstrate the effectiveness of our testing methodology in assessing both functional correctness and extra-functional properties. Our framework implements 17 distinct metrics that encompass syntactic analysis, semantic evaluation, and behavioral evaluation through LLM judges. The study reveals significant information about how different architectural choices and parameter configurations affect system performance, particularly highlighting the impact of temperature and top-p parameters on response quality. The tests were carried out on a tourism recommendation system for the Värmland region, utilizing standard and RAG-enhanced configurations. The results indicate that the newer LLM versions show modest improvements in performance metrics, though the differences are more pronounced in response length and complexity rather than in semantic quality. The research contributes practical insights for implementing robust testing practices in LLM-RAG systems, providing valuable guidance to organizations deploying these architectures in production environments.
 
\end{abstract}

\begin{IEEEkeywords}
Large Language Models (LLM), Software Quality Testing, Retrieval-Augmented Generation (RAG), Extra-Functional Properties, ML System Testing, AI Quality Assurance

\end{IEEEkeywords}

\section{Introduction}

The rapid advancement of Large Language Models (LLMs) has revolutionized how we develop software applications, particularly in domains requiring sophisticated natural language understanding and generation capabilities \cite{Brown2020}. These AI-powered systems have emerged as transformative tools in creating intelligent applications that can process, understand, and generate human-like responses. For example, in the tourism sector, LLM-based applications have become increasingly prevalent, offering personalized recommendations, answering queries about local attractions, and providing cultural and historical information to visitors. These systems often employ Retrieval-Augmented Generation (RAG) architectures to combine the powerful language capabilities of LLMs with accurate, up-to-date domain-specific information \cite{Lewis2020}.

The integration of LLMs into software systems brings unprecedented challenges in quality assurance and testing. Unlike traditional software systems with deterministic behaviors, LLM-based applications exhibit characteristics that make conventional testing approaches insufficient. These systems demonstrate non-deterministic outputs, context-dependent behaviors, and potential hallucinations that can impact their reliability and trustworthiness \cite{WangTSE2024}. The complexity increases when considering RAG systems, where the quality of responses depends not only on the LLM's capabilities but also on the accuracy and relevance of the retrieved information \cite{zhao2024arxive}. Testing such systems requires new methodologies that can effectively evaluate both functional correctness and extra-functional properties such as reliability, consistency, and robustness \cite{hudson2024softwareeng}.

Recent advances in LLM evaluation, as documented by Chang et al. \cite{ChangSurvey2024}, have primarily focused on functional aspects such as response accuracy and generation quality. However, the integration of these models into production systems necessitates a broader evaluation scope that encompasses system qualities beyond mere functional correctness. This expanded scope becomes particularly critical when examining the interaction between retrieval and generation components in RAG systems, where traditional testing approaches prove insufficient for ensuring system reliability and performance under varying operational conditions.


The testing landscape for LLM-RAG systems is further complicated by their hybrid nature. The retrieval component must maintain high accuracy and performance while dealing with potentially large knowledge bases, while the generation component needs to produce consistent, high-quality outputs based on the retrieved context. Wang et al. \cite{wang2023measuring} demonstrate how semantic consistency testing can reveal underlying system behaviors that affect overall reliability. Their work highlights the need for systematic approaches to evaluate both component-level and system-level properties.

Recent research has made significant strides in understanding these challenges. Es et al. \cite{Xuanfan2024} present a systematic evaluation framework that emphasizes the importance of comprehensive testing strategies. Furthermore, Dhuliawala et al. \cite{dhuliawala2024} introduce verification approaches that address the critical issue of hallucination in LLM outputs, a concern that becomes even more pronounced in RAG architectures where factual accuracy is paramount.

Our research builds upon these foundations by implementing and validating a comprehensive testing approach for LLM-RAG systems. Through systematic empirical evaluation, we demonstrate how different architectural choices and parameter configurations affect various system properties. Our implementation encompasses multiple testing dimensions, including semantic similarity assessment, response consistency evaluation, and systematic analysis of system behavior under varying conditions. By adapting existing testing methodologies and metrics to the specific challenges of RAG architectures, we provide practical insights into effective quality assurance practices for these complex systems. The main contributions of our work include:

\begin{itemize}
\item A systematic implementation of testing methodologies specifically adapted for LLM-RAG systems, demonstrated through detailed empirical evaluation
\item Comprehensive analysis of how different system configurations and parameters affect various quality attributes in RAG architectures
\item Practical insights and guidelines for implementing effective testing practices in LLM-RAG systems, derived from extensive experimental results
\item Validation of our testing approach through a real-world case study, providing concrete evidence of its effectiveness in production environments
\end{itemize}

These contributions advance our understanding of how to effectively test and ensure quality in LLM-RAG systems, with implications extending beyond our specific case study to the broader domain of AI-augmented software systems. Our work is particularly relevant as organizations increasingly adopt LLM-RAG architectures in production environments. The insights and implementation strategies we present provide valuable guidance for practitioners seeking to establish robust testing practices for their AI-augmented applications. Furthermore, our findings contribute to the broader discourse on quality assurance in complex software systems, offering practical approaches to addressing the challenges that arise when integrating advanced AI capabilities into production applications.


\section{Background and Related Work}

The testing and evaluation of LLMs have emerged as a critical research focus as these systems become increasingly integrated into production applications. Chang et al. \cite{ChangSurvey2024} provides a comprehensive framework to understand the current state of LLM evaluation in their systematic survey. Their work synthesizes existing evaluation approaches and metrics, highlighting that effective LLM assessment requires considering multiple dimensions, including accuracy, reliability, and robustness. This multi-faceted approach to evaluation has particular relevance to our work on testing RAG systems in tourism applications, where both response quality and factual accuracy are crucial.

In the specific context of RAG systems, Es et al. \cite{Shahul2024} introduced RAGAS, a pioneering framework for automated evaluation. Their work establishes standardized metrics to assess critical aspects of RAG systems, including faithfulness to source documents, context relevance, and quality of the answers. The RAGAS framework has particular significance for our testing methodology, as it provides systematic ways to evaluate how effectively our tourism RAG system retrieves and utilizes domain-specific information while maintaining accuracy in its recommendations.


The challenge of ensuring reliable LLM output has been comprehensively addressed by Wang et al. \cite{wang2023measuring}, who developed methods to measure semantic consistency in LLM responses. Their research demonstrates how LLMs can produce inconsistent output even with similar inputs, a finding that directly influenced our approach to testing response consistency in tourism recommendations. Their proposed semantic consistency evaluation metrics provide valuable tools for assessing the reliability of LLM-generated travel advice and ensuring consistent user experiences.

Recent work by Ni and Li \cite{Xuanfan2024} presents a systematic evaluation framework specifically focused on natural language generation tasks. Their comprehensive assessment methodology examines multiple aspects of the generated text, including fluency, coherence, and task-specific effectiveness. This work has informed our approach to evaluating the quality of generated travel recommendations, particularly in assessing how well our system maintains natural language quality while providing accurate tourism information. A significant advance in treating LLM hallucination comes from Dhuliawala et al. \cite{dhuliawala2024}, who introduced the chain of verification approach. Their method provides robust techniques for detecting and reducing hallucination in LLM outputs, particularly crucial for RAG systems where factual accuracy is paramount. We have incorporated aspects of their verification methodology into our testing framework to ensure the accuracy of the retrieved information and the travel recommendations generated.

Our work builds on these foundations while addressing the challenges of testing RAG systems in tourism applications. We extend existing frameworks by incorporating domain-specific considerations and developing new metrics for evaluating the quality of travel recommendations. In particular, we adapt the RAGAS evaluation metrics and the consistency testing approaches of Wang et al. to create a comprehensive testing framework specifically tailored to tourism-focused RAG systems. We used Evidently\footnote{Evidently Website https://www.evidentlyai.com/} framework, an open source tool, to implement these testing strategies. This integration of existing methodologies with domain-specific requirements allows us to effectively assess both the technical performance and practical utility of our system.


\section{Architecture of the Application Under Test}

This section outlines the RAG framework employed for enhancing location-based recommendations in a travel planning application used for testing. The system architecture, illustrated in Figure \ref{fig:architecture}, implements a comprehensive pipeline that transforms raw tourism data into contextually relevant travel recommendations through four main stages: data acquisition and preprocessing, embedding generation, retrieval mechanism, and augmented query generation.

The architectural diagram in Figure \ref{fig:architecture} represents the complete data flow from initial collection to final generation of the travel plan. The system's components are arranged in a logical sequence, with each stage building upon the outputs of previous processes. 

\subsection{Data Acquisition}
Tourism-related data was sourced from the V\"{a}rmland Tourism API\footnote{https://turid.visitvarmland.com/api/v8/docs}, an official repository providing extensive information on regional attractions, events, and services. As shown in the leftmost section of Figure \ref{fig:architecture}, the data collection process begins with the Turid website, which serves as the primary data source. The API endpoints were queried to extract structured data encompassing titles, descriptions, categories, municipalities, and metadata such as operating hours and event schedules.

The initial data collection component interfaces directly with the tourism database to gather comprehensive information. This data is then transformed into a JSONL format, which standardizes the information for subsequent processing stages. The dataset reflects the diversity of offerings in the region, ensuring inclusivity of interests such as nature, literature, design, and science.


\subsection{Embedding Generation and Processing}
The system implements an embedding generation pipeline, depicted in the central portion of Figure \ref{fig:architecture}. The data processing stage employs parallel processing capabilities to handle large volumes of information efficiently. This architectural decision was crucial for maintaining the system's performance during testing, as it allowed simultaneous processing of multiple data streams.

The embedding generation component transforms the processed tourism data into high-dimensional vector representations, enabling semantic search capabilities. This transformation occurs through a parallel processing framework that optimizes the utilization of computational resources and reduces processing latency during testing operations.

\subsection{Similarity Indexing and Storage}

The generated embeddings are organized within a FAISS (Facebook AI Similarity Search) database \cite{douze2024faiss}, labeled "FAISS database" on the diagram. This component provides efficient similarity search capabilities and serves as the primary retrieval mechanism for the system. The implementation of FAISS represents a critical architectural decision that directly influences both the accuracy of the retrieval and the response times of the system during the testing procedures.


\subsection{Query Processing and Travel Plan Generation}

\begin{figure}
    \centering
    \includegraphics[width=0.9\linewidth]{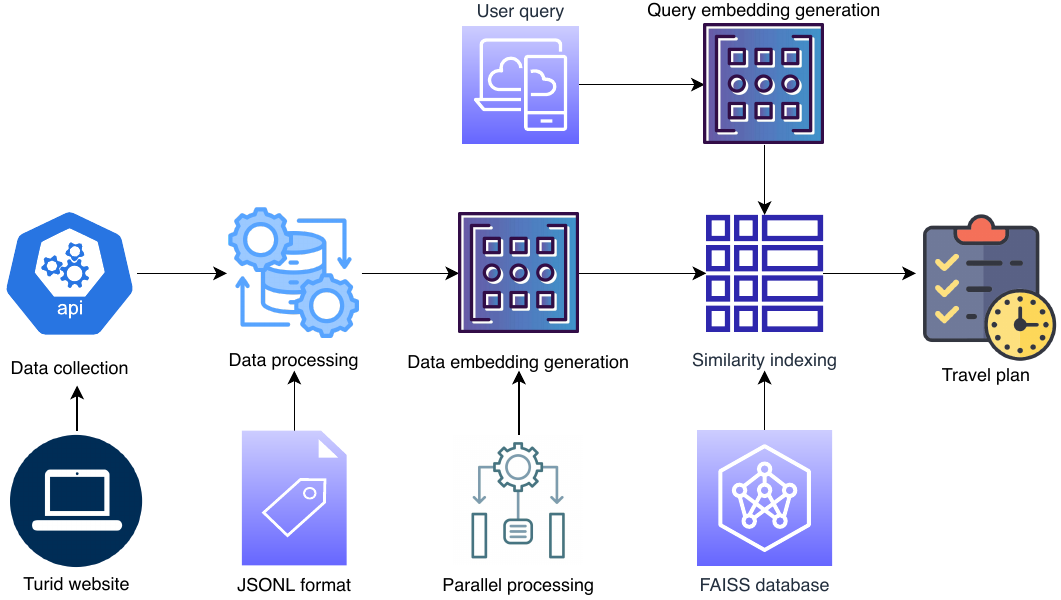}
    \caption{Architecture of the RAG-based tourism recommendation system showing the data flow from collection through travel plan generation.}
    \label{fig:architecture}
\end{figure}

The final stage of the architecture handles user queries and travel plan generation. As illustrated in the upper part of Figure \ref{fig:architecture}, user queries undergo a parallel embedding generation similar to the initial data processing stage. The query embedding is then used to perform similarity searches against the FAISS index, retrieving contextually relevant tourism information.

The travel plan generation component combines the retrieved context with the original query to produce coherent travel recommendations. This component represents the culmination of the RAG pipeline, where the system's ability to maintain semantic coherence and generate relevant recommendations was extensively tested.

\section{Testing Methodology}

This section provides a detailed overview of the objectives, scope, implementation, execution, and metrics of the tests conducted within this paper. 

\subsection{Testing Objectives}

The main objective of this study was to create an automated set of tests that can be performed on various models and configurations to track and evaluate the performance of an LLM application throughout its lifecycle. To illustrate this, tests were performed on a travel planning application with varying parameters to simulate regression tests in a real software development context.

The first goal of these tests was to identify any significant differences between the different LLMs used within the application. The second goal was to evaluate how changes in model parameters for response generation affect the test results of each model. 

Furthermore, while accurate travel suggestions for specific regions can only be ensured with a RAG system, the third goal was to assess whether the use of RAG influences the test results. This was achieved by comparing the evaluation of responses generated with the RAG system and those formulated without the utilization of additional context-specific data.

\subsection{Scope of Testing}

Our comprehensive testing framework encompasses 17 distinct tests organized into three fundamental categories: text metrics, semantic similarity assessments, and LLM-based evaluations. Each category serves a specific purpose in assessing different aspects of the generated responses.

The foundational layer consists of five general text metrics that evaluate the structural characteristics of responses. These metrics analyze both surface-level features (character count, word count, sentence count) and more nuanced aspects like the proportion of non-letter characters and out-of-vocabulary terms. We employ the Natural Language Toolkit's vocabulary database for out-of-vocabulary detection, which helps identify potential comprehension barriers\footnote{https://github.com/evidentlyai/evidently/blob/main/docs/book/reference/all-metrics.md}. While these metrics don't directly measure response quality, they serve as essential indicators of structural consistency and enable systematic tracking of changes throughout the development lifecycle.

Building upon the basic metrics, our framework implements four semantic similarity tests using two complementary methodologies. The first approach utilizes word embedding vectors, leveraging their high-dimensional semantic space representation where similar meanings cluster together. This method calculates cosine similarity between vector representations to measure semantic alignment\footnote{https://github.com/evidentlyai/evidently/blob/main/docs/book/reference/all-metrics.md}. We apply this in two contexts: comparing responses to their original prompts (measuring request adherence) and comparing responses to reference examples (measuring adherence to expected quality standards).

The second semantic evaluation methodology employs BERT encoding, which offers a more sophisticated approach to capturing semantic content through its transformer-based architecture and unsupervised learning capabilities. Similar to the embedding vector approach, we implement two BERT-based tests that evaluate prompt-response and response-reference similarities, providing a deeper understanding of semantic relationships.


The most sophisticated layer of our testing framework consists of eight LLM-based evaluation tests that act as automated judges. Three of these tests provide continuous scale evaluations:

\begin{itemize}
    \item Sentiment analysis (scale: -1 to 1)
    \item Toxicity detection (scale: 0 to 1)
    \item Neutrality assessment (scale: 0 to 1)
\end{itemize}

The remaining five tests perform categorical assessments with binary classifications:

\begin{itemize}
    \item Request fulfillment validation (DECLINE/OK)
    \item Privacy compliance verification (PII/OK)
    \item Tone analysis (POSITIVE/NEGATIVE)
    \item Bias detection (BIAS/OK)
    \item Content safety assessment (TOXICITY/OK)
\end{itemize}




\subsection{Implementation Architecture}

The entire testing framework is implemented using Evidently, an open-source platform designed for machine learning evaluation. Our implementation prioritizes automation and reproducibility, requiring minimal user intervention. The system accepts three key input parameters:

\begin{itemize}
    \item LLM model selection
    \item Model parameter configuration
    \item RAG integration toggle
\end{itemize}

Each test execution follows a systematic process:

\begin{enumerate}
    \item Response generation across all test cases
    \item Application of the 17-metric evaluation suite
    \item Result storage in a structured database
    \item Automated generation of individual test graphs
    \item Creation of regression analysis visualizations
\end{enumerate}

\begin{figure}
   \centering
   \includegraphics[width=\linewidth]{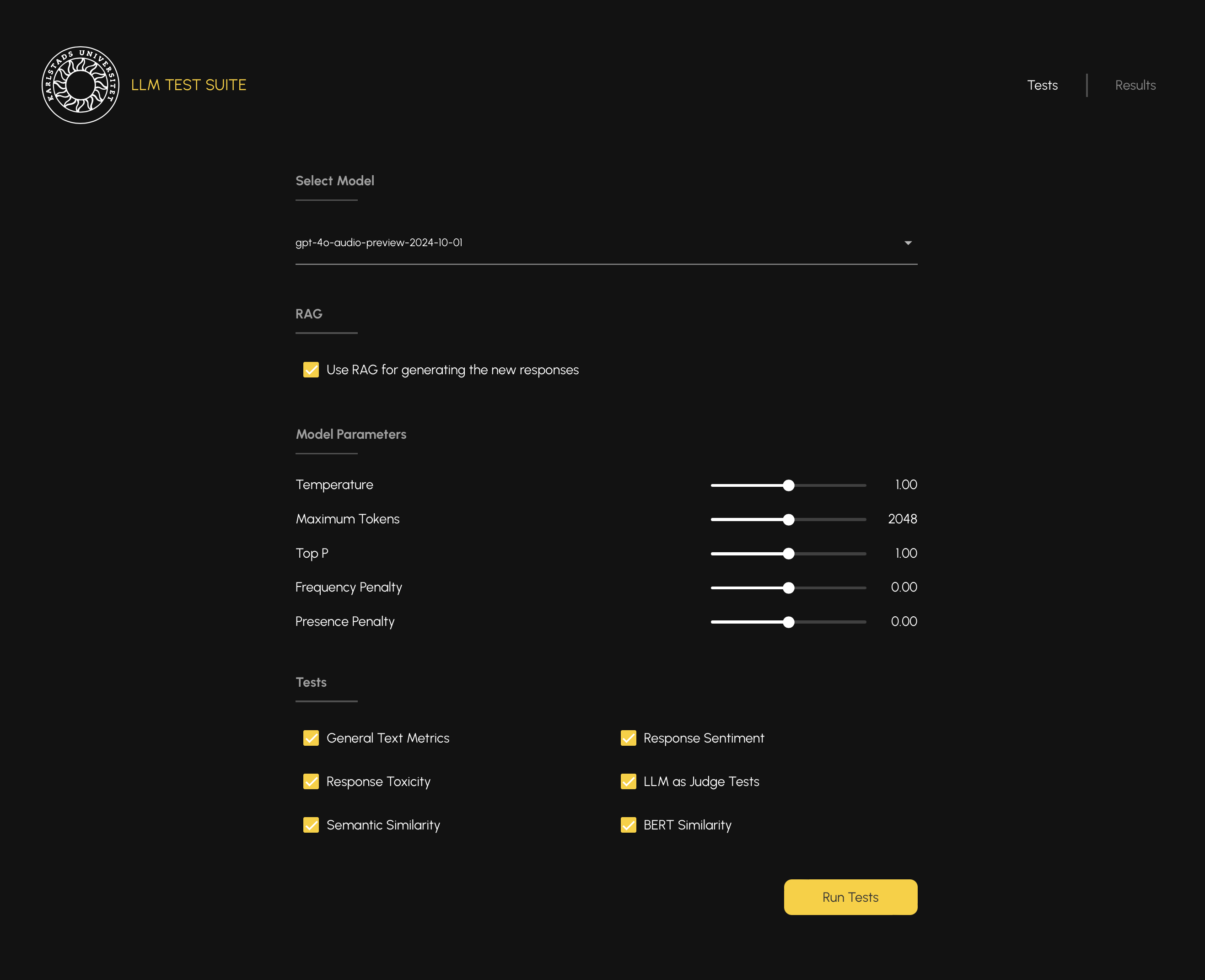}
   \caption{The Graphical User Interface of the Testing Suite}
   \label{fig:test_suite_gui}
\end{figure}

The framework supports both graphical and command-line interfaces, facilitating integration into existing development pipelines. This dual-interface approach enables both interactive exploration through the GUI (shown in Figure \ref{fig:test_suite_gui}) and automated testing through command-line operations. The system's ability to execute multiple test runs with varying parameters supports comprehensive evaluation of different model configurations.

\subsection{Test Case Design and Coverage}

Our evaluation methodology is founded on a systematically designed test suite comprising 10 comprehensive test cases for assessing the travel recommendation system's capabilities in the Värmland region context. Each test case is structured as a paired set: a carefully crafted prompt that authentically simulates user queries, and a corresponding reference response that establishes quality benchmarks for both content and structural characteristics.

The reference responses were developed through rigorous analysis of tourism domain requirements, incorporating both factual accuracy and natural language quality standards. While these references serve as quality benchmarks, our evaluation framework acknowledges the inherent diversity in valid travel recommendations, recognizing that multiple response variations can be equally valid for a given query.

To ensure comprehensive system evaluation, our test cases were designed using a systematic coverage matrix that addresses multiple dimensions of system functionality:

\begin{itemize}
   \item Query complexity variations, ranging from simple point-of-interest requests to multi-faceted travel planning
   \item Temporal and seasonal considerations specific to Värmland tourism
   \item Edge cases testing system robustness and boundary conditions
   \item Uncommon request patterns that challenge system adaptability
\end{itemize}

The selection of 10 test cases represents an optimized balance between evaluation thoroughness and computational efficiency, determined through preliminary testing that indicated diminishing returns in coverage beyond this number. During execution, each test case processes three key components: the query prompt, model-specific parameters, and when applicable, the RAG fact file for knowledge augmentation.




\subsection{Data Collection and Management Framework}

Our data collection architecture implements a sophisticated pipeline for gathering, storing, and analyzing test results. The framework employs a relational database structure that maintains comprehensive test execution records through the following organization:

\begin{itemize}
   \item Primary test execution records linking model identifiers with parameter configurations
   \item Generated response storage with full context preservation
   \item Metric computation results across all 17 evaluation dimensions
   \item System performance metadata for execution analysis
   \item Evidently-generated HTML reports for visual analysis and presentation
\end{itemize}

This architecture enables both real-time analysis and longitudinal performance studies while maintaining complete traceability of system behavior across different configurations.

\section{Experimental Methodology}

Our experimental evaluation framework implemented a comprehensive assessment protocol comprising 24 distinct test configurations. The evaluation matrix incorporated three leading LLM variants: GPT 3.5 Turbo, GPT 4o, and GPT 4o Mini, each tested across multiple parameter configurations and RAG integration scenarios.

The parameter space exploration focused on two critical response generation controls:

\begin{itemize}
   \item Temperature parameter: Controls stochastic variation in response generation
   \item Top-p parameter: Governs diversity in the output token selection process
\end{itemize}

For each model, we established a testing matrix with the parameter configurations as shown in Table \ref{tab:param_config}.

\begin{table}
\centering
\caption{Parameter Configurations for Model Testing}
\begin{tabular}{|l|c|c|}
\hline
Configuration & Temperature & Top-P \\
\hline
Baseline & 0 & 0 \\
Diverse & 1 & 0 \\
Controlled & 0 & 2 \\
Experimental & 1 & 2 \\
\hline
\end{tabular}

\label{tab:param_config}
\end{table}

Additional model parameters, including maximum token length, frequency penalty, and presence penalty, were maintained at their default values based on preliminary testing that indicated minimal impact on response quality in our specific use case\footnote{https://platform.openai.com/docs/api-reference/chat/create}.

The experimental protocol implemented a full factorial design:
\begin{itemize}
   \item Three distinct model architectures
   \item Four parameter configurations per model
   \item Dual testing modes (with and without RAG integration)
\end{itemize}

This systematic approach yielded a comprehensive evaluation matrix of 24 distinct test configurations (3 × 4 × 2), enabling detailed analysis of model behavior across different operational scenarios. While fact-verification metrics were intentionally excluded due to their intrinsic dependency on RAG integration, this design choice acknowledged the anticipated performance differential between RAG and non-RAG configurations in domain-specific knowledge accuracy.


\section{Results and Analysis}

\subsection{Results}

The data collected from the test runs allows a comparison of how the three LLMs perform in each of the tests with different model parameters, as well as providing some insight into whether there is a difference in quality for responses generated with or without RAG. 

\begin{figure}
    \centering
    \includegraphics[width=\linewidth]{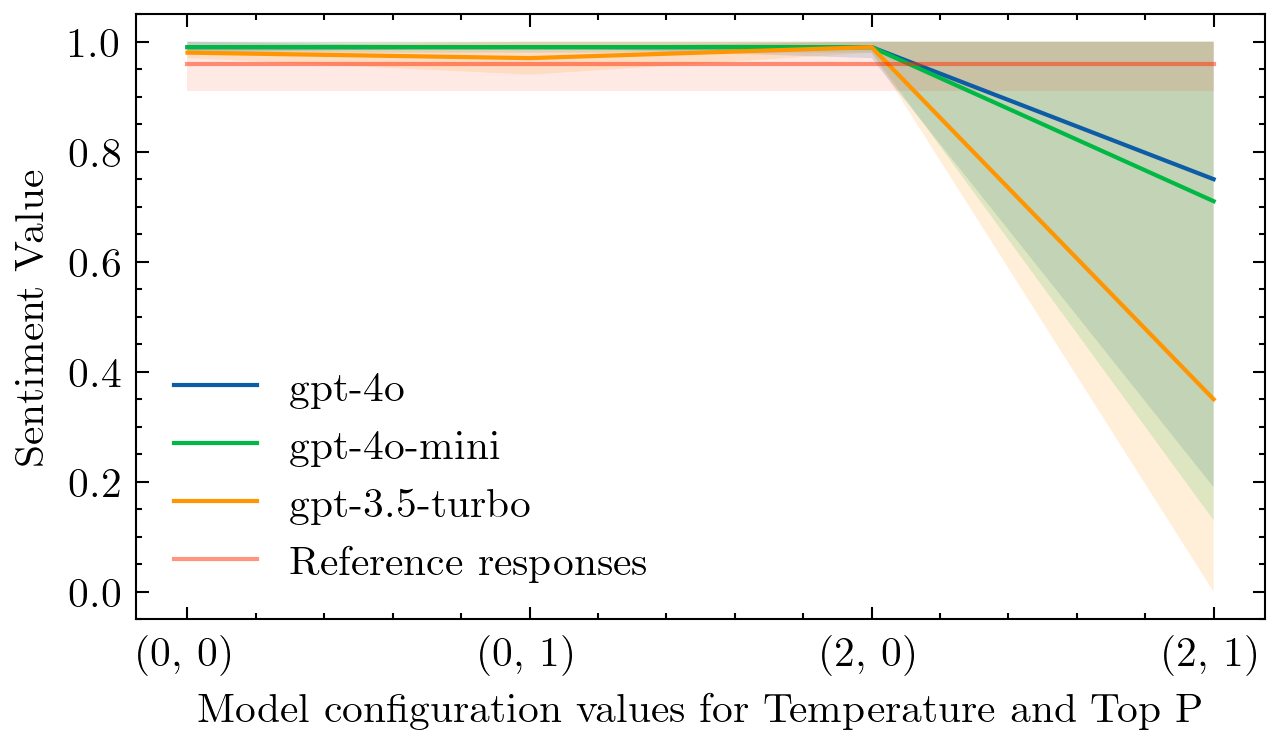}
    \caption{Response Sentiment for new RAG responses}
    \label{fig:sentiment_with_rag}
\end{figure}

One notable characteristic that can be seen across the results from all tests is a strong decrease in model performance with extreme values for the two parameters “Temperature” and “Top P”. With both of these parameters at their maximum value the test results show a steep decline of up to 64\% compared to the other parameter configurations, dropping from a mean sentiment estimation of 0.99 to 0.35 across all test cases for the model “GPT 3.5 Turbo” (Figure \ref{fig:sentiment_with_rag}). However, due to the low number of sampled parameter values, it is difficult to tell at which threshold specifically this drop in performance occurs.

\begin{figure}
    \centering
    \includegraphics[width=\linewidth]{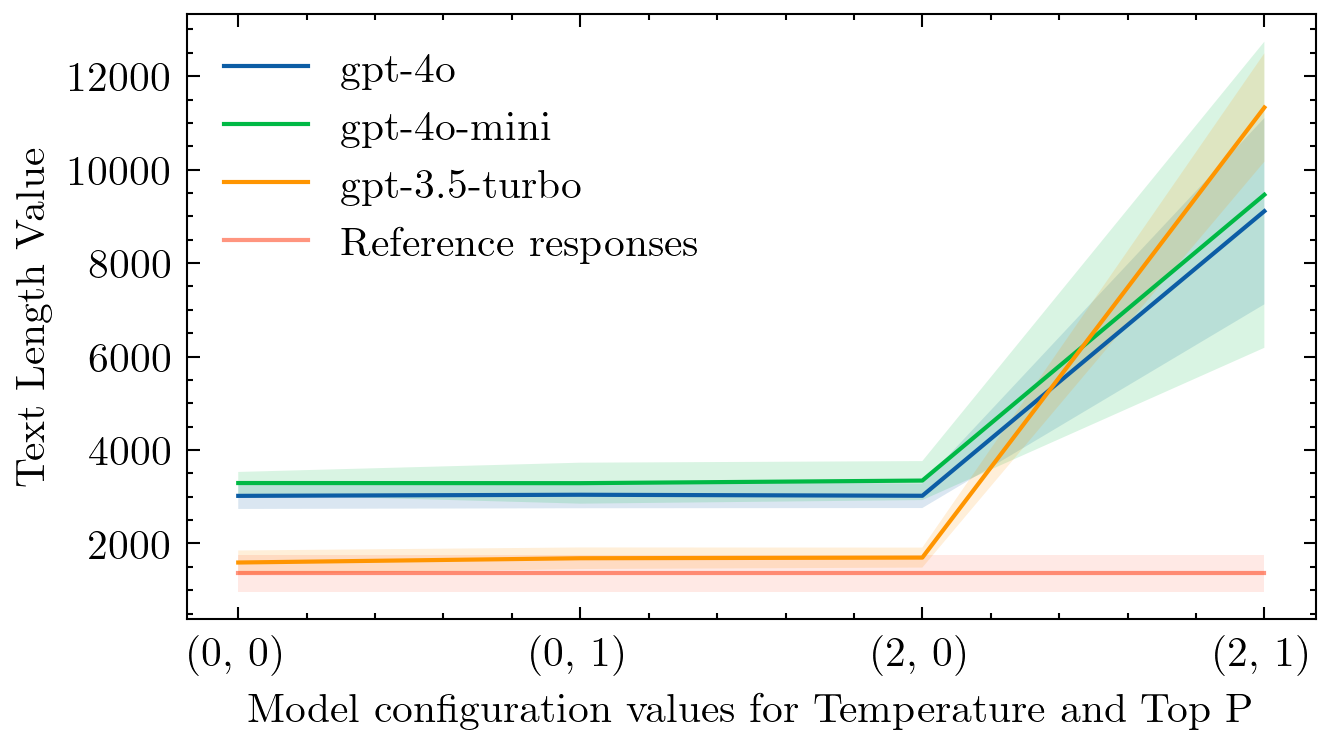}
    \caption{Response Length for new responses, created without RAG}
    \label{fig:text_length_no_rag}
\end{figure}

Overall the three model variations, GPT 3.5 Turbo, GPT 4o and GPT 4o Mini, show very similar results across all tests and parameter configurations. The only exception from that can be found in the text metrics such as text length and sentence count. While GPT 4o and GPT 4o Mini perform similar, even in these tests, GPT 3.5 Turbo produces significantly shorter responses as seen in Figure \ref{fig:text_length_no_rag}, while also using fewer non-letter characters.

\begin{figure}
    \centering
    \includegraphics[width=\linewidth]{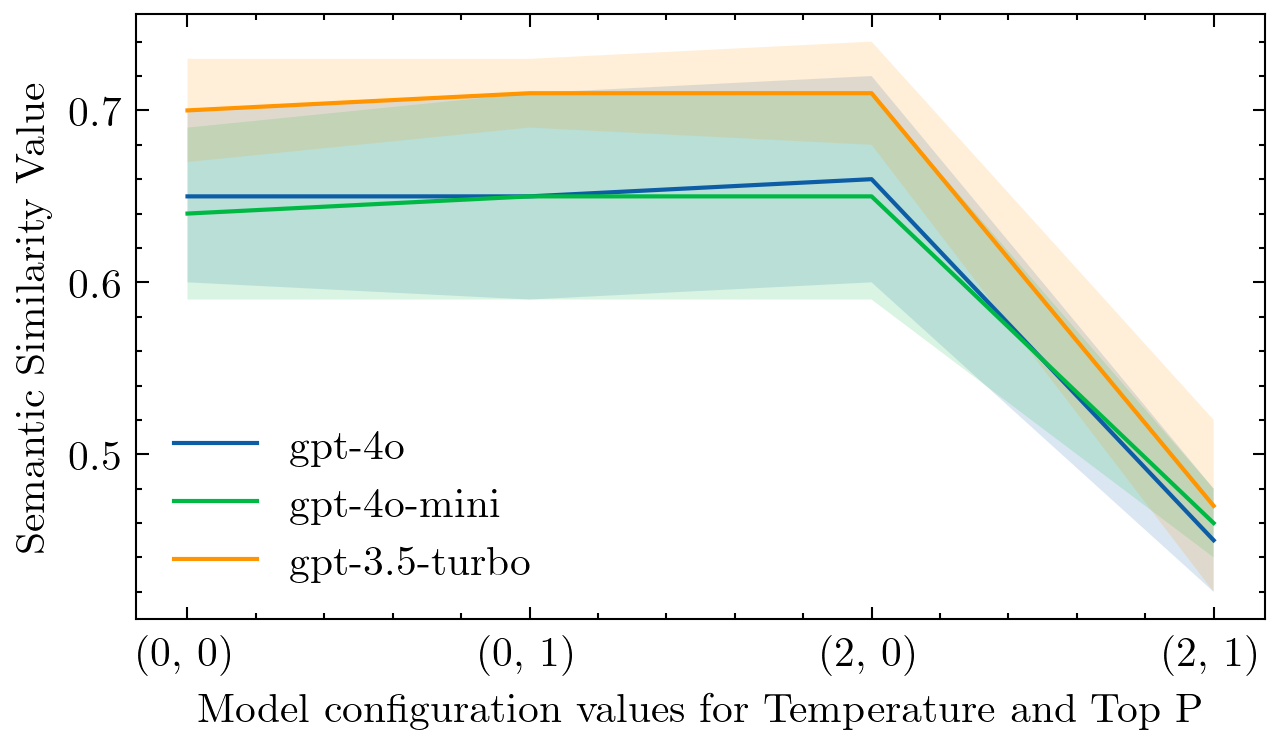}
    \caption{Semantic Similarity between new RAG responses and reference responses evaluated using BERT}
    \label{fig:semantic_similarity_bert_rag}
\end{figure}

\begin{figure}
    \centering
    \includegraphics[width=\linewidth]{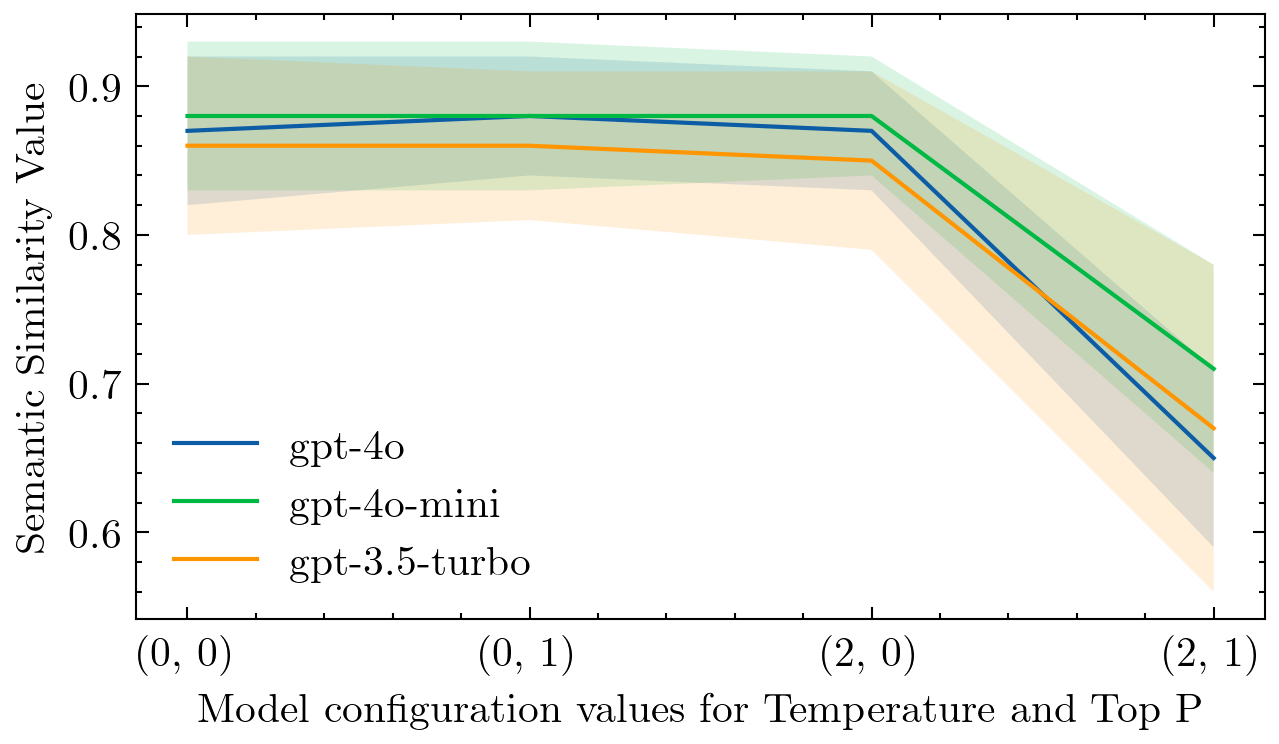}
    \caption{Semantic Similarity between new RAG responses and reference responses evaluated using embeddings}
    \label{fig:semantic_similarity_rag}
\end{figure}

When comparing semantic similarity between the new responses and the reference responses, GPT 3.5 Turbo slightly outperforms the other two models in the BERT similarity test as seen in Figure \ref{fig:semantic_similarity_bert_rag}. However, the difference is relatively small and is not replicated by the second semantic similarity test, which is based on word embeddings (Figure \ref{fig:semantic_similarity_rag}). Overall, the semantic similarity test using BERT shows a lower similarity score across all models, both between the new response and the reference response, as well as between the new response and the test prompt, compared to the semantic similarity score based on word embeddings. This is possibly due to the two tests having different internal scalings.

\begin{figure}
    \centering
    \includegraphics[width=\linewidth]{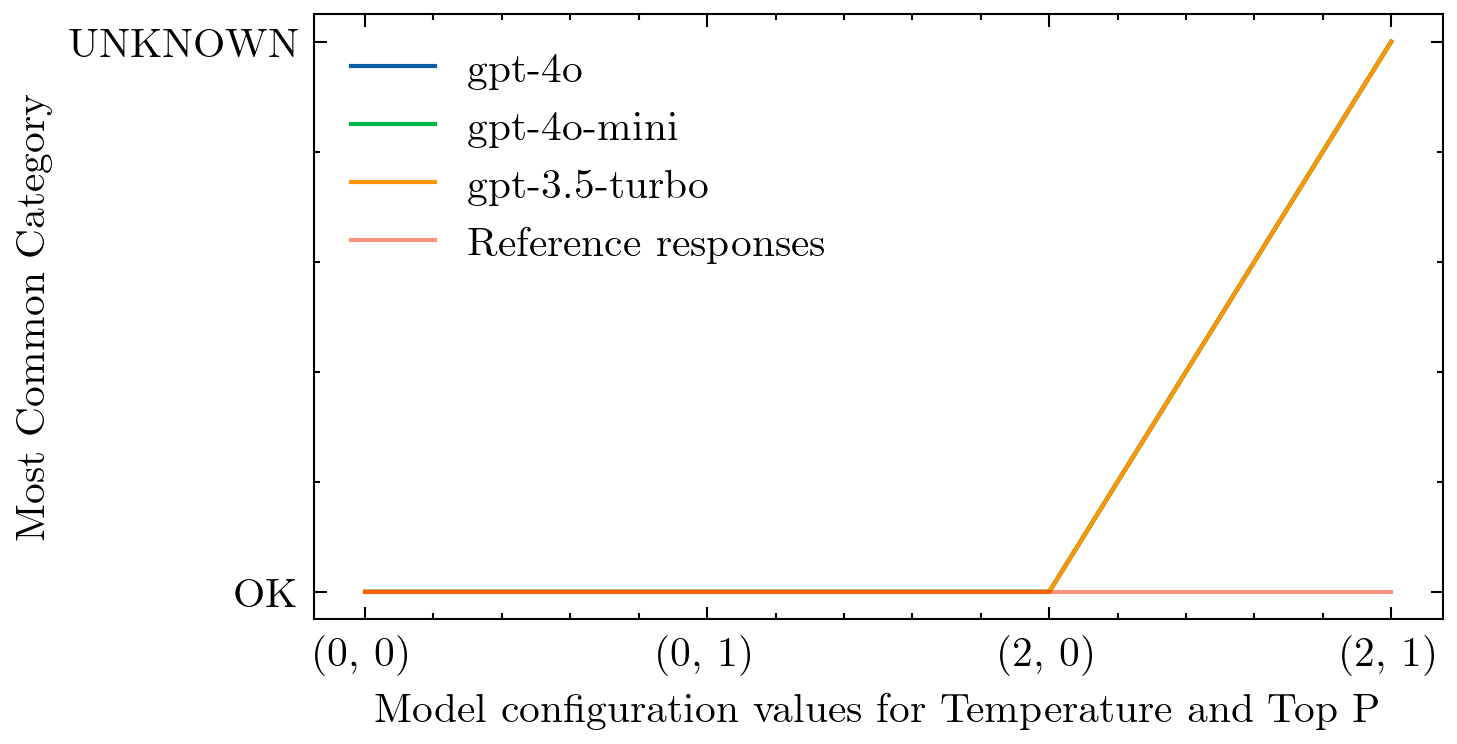}
    \caption{Response Toxicity judged using an LLM for new responses, generated without RAG}
    \label{fig:toxicity_llm_no_rag}
\end{figure}

The results of the toxicity tests show unanimously that the responses from all models don’t contain any toxic contents, unless the model parameters reach their highest values which introduce the most randomness into the model’s results. However, even for this parameter configuration, the toxicity test using an LLM as a judge only assigns the category “unknown” to the new responses, not “toxic” as seen in Figure \ref{fig:toxicity_llm_no_rag}. Furthermore, the score-based toxicity test increases from 0.0 for the other parameter configurations to a value around 0.04.

\begin{figure}
    \centering
    \includegraphics[width=\linewidth]{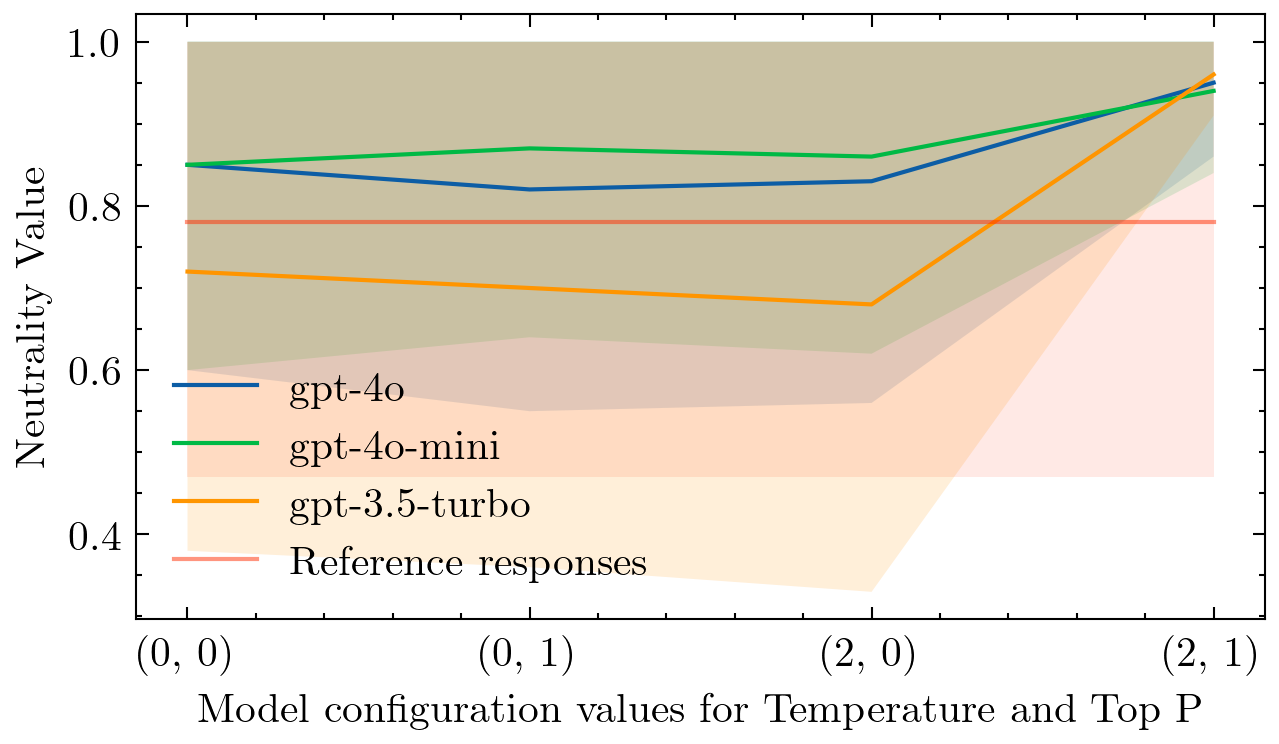}
    \caption{Response Neutrality of new RAG responses}
    \label{fig:response_neutrality_rag}
\end{figure}

In the category of response neutrality the two newer models, GPT 4o and GPT 4o Mini, outperform the older model, especially when using RAG (Figure \ref{fig:response_neutrality_rag}). The test has a really high standard deviation across all models, indicating the neutrality metric is quite unstable and probably requires a larger set of test cases to provide a more accurate result. While the neutrality results remain consistent across most of the parameter configurations, they increase significantly for the most extreme configuration. This suggests that a high neutrality value does not necessarily equate to a good response.

The other tests such as response sentiment, response bias, response denial, response negativity and the test for personal identifiable information all showed no difference between the tested models, parameter configurations and whether or not RAG was used. Only the most extreme parameter configuration consistently shows a degradation in performance, as mentioned above.

\subsection{Analysis}

Our comprehensive evaluation reveals several key findings regarding the performance characteristics of different LLM versions and their configurations in tourism applications. The newer models (GPT-4o and GPT-4o Mini) demonstrate incremental improvements in our test metrics compared to GPT-3.5 Turbo, although these improvements manifest primarily in response structure rather than fundamental quality measures. The most notable improvement appears in response length and complexity, where the newer models consistently generate more detailed travel recommendations. This increased detail, while potentially beneficial for comprehensive travel planning, should be considered within the specific application context and user experience requirements.

An interesting observation emerges regarding the impact of the RAG integration on model performance. While our general quality metrics showed minimal variation between RAG and non-RAG configurations, this finding requires careful interpretation. The RAG architecture proves particularly valuable in ensuring factual accuracy and regional specificity in travel recommendations, aspects that extend beyond our general quality metrics. This suggests that traditional evaluation metrics might need adaptation to fully capture the benefits of RAG in domain-specific applications.

Parameter configuration analysis revealed critical insights into model behavior. Most notably, conservative parameter settings (lower temperature and top-p values) consistently produced more reliable and coherent responses. However, when these parameters reached extreme values (temperature of 2.0 and top-p of 1.0), response quality degraded significantly, often resulting in incoherent or contextually inappropriate suggestions. This finding establishes clear operational boundaries for deploying these models in production environments.

The neutrality metric exhibited particularly interesting behavior, showing considerable variation across test cases. While this variation might initially appear concerning, deeper analysis revealed that it did not correlate with degradation in other quality metrics or overall response utility. This suggests that response neutrality, while worth monitoring, may not be a critical factor in travel recommendation applications where some degree of opinion or preference expression is expected and potentially beneficial.
All tested models, when operating within their optimal parameter ranges, demonstrated robust performance across our evaluation metrics. This consistency suggests that the choice between model versions might be better guided by specific application requirements, such as response length preferences or computational resource constraints, rather than fundamental quality differences. The RAG architecture, while it does not significantly affect our general quality metrics, is essential to maintain domain-specific precision and relevance in travel recommendations.
These findings provide valuable information for organizations that implement LLM-RAG systems, particularly in tourism applications, while also highlighting areas where evaluation methodologies might be improved to better capture domain-specific performance characteristics.

\section{Threats to Validity}

Several considerations regarding the validity of our findings warrant discussion, although our testing methodology demonstrates robust evaluation capabilities within the specified context. Our study focused on tourism applications using three state-of-the-art LLM variants from OpenAI, chosen for their widespread adoption and documented reliability in production environments. While this scope enabled detailed comparative analysis, future work could benefit from including models from additional providers to enhance generalizability.

The parameter space exploration utilized four distinct configurations chosen to represent key operational points in the models' behavior spectrum. This strategic sampling approach, while not exhaustive, provided clear insights into performance patterns and critical transition points. Our selection of these specific configurations was guided by common production deployment scenarios and resource optimization considerations.

Our test suite comprised ten carefully crafted test cases, designed to cover diverse aspects of travel recommendation scenarios. This number was determined through a balance of comprehensive coverage and resource efficiency, particularly considering the computational demands of large-scale LLM testing. The test cases were specifically engineered to exercise different aspects of the system's capabilities, from basic information retrieval to complex multi-faceted travel planning scenarios.

The focus on travel recommendation systems provided a well-defined domain for evaluating RAG capabilities, particularly given the structured nature of tourism data and clear evaluation criteria. While this specific application context might limit direct generalization to other domains, the underlying testing methodology and framework design principles remain applicable across different LLM-RAG implementations.

To address these validity considerations, our study employed rigorous testing protocols and comprehensive measurement approaches. The combination of syntactic, semantic, and behavioral metrics provides multiple perspectives on system performance, helping to mitigate the limitations of any single evaluation approach. Furthermore, the consistency of our findings across different parameter configurations and test cases suggests the robustness of our conclusions within the studied context.



\section{Conclusion and Future Work}

This research presents a systematic approach to testing LLM-RAG systems through a comprehensive evaluation framework demonstrated in a tourism recommendation system. Our empirical results reveal that while newer LLM versions show improvements over their predecessors, these enhancements primarily manifest in response length and complexity rather than fundamental quality metrics. The testing framework highlighted the critical importance of parameter configuration, with extreme values for temperature and top-p parameters consistently leading to significant degradation in response quality.

A key finding was that while RAG integration did not substantially affect the general quality metrics we tested, it proved crucial for maintaining factual accuracy in domain-specific responses. This observation emphasizes the need for specialized testing approaches that can effectively evaluate both general language capabilities and domain-specific knowledge integration in RAG systems. The research also identified limitations in current testing methodologies, particularly in evaluating semantic consistency and response neutrality across different operational contexts.

Looking forward, this work lays the groundwork for future research directions, focusing on expanding the evaluation framework to include a broader range of LLM architectures, developing more sophisticated metrics for domain-specific knowledge integration, and investigating parameter configurations with finer granularity. The findings and methodologies presented provide a foundation for developing more robust testing practices for LLM-RAG systems, contributing to the broader discourse on quality assurance in AI-augmented software systems.

\section*{Acknowledgment}
This research was conducted as part of the PERUPP project within the Academy for Smart Specialisation, a strategic partnership between Karlstad University and Region Värmland. The project was funded by the European Regional Development Fund through the VisitorsXLab project.

\balance

\bibliographystyle{ieeetr}

\bibliography{ReferencesICST}

\begin{thebibliography}{10}

\bibitem{Brown2020}
T.~B. Brown, B.~Mann, N.~Ryder, M.~Subbiah, J.~D. Kaplan, P.~Dhariwal, A.~Neelakantan, P.~Shyam, G.~Sastry, A.~Askell, S.~Agarwal, A.~Herbert-Voss, G.~Krueger, T.~Henighan, R.~Child, A.~Ramesh, D.~M. Ziegler, J.~Wu, C.~Winter, C.~Hesse, M.~Chen, E.~Sigler, M.~Litwin, S.~Gray, B.~Chess, J.~Clark, C.~Berner, S.~McCandlish, A.~Radford, I.~Sutskever, and D.~Amodei, ``Language models are few-shot learners,'' in {\em Advances in Neural Information Processing Systems} (H.~Larochelle, M.~Ranzato, R.~Hadsell, M.~Balcan, and H.~Lin, eds.), vol.~33, pp.~1877--1901, Curran Associates, Inc., 2020.

\bibitem{Lewis2020}
P.~Lewis, E.~Perez, A.~Piktus, F.~Petroni, V.~Karpukhin, N.~Goyal, H.~K\"{u}ttler, M.~Lewis, W.-t. Yih, T.~Rockt\"{a}schel, S.~Riedel, and D.~Kiela, ``Retrieval-augmented generation for knowledge-intensive nlp tasks,'' in {\em Proceedings of the 34th International Conference on Neural Information Processing Systems}, NIPS '20, (Red Hook, NY, USA), Curran Associates Inc., 2020.

\bibitem{WangTSE2024}
J.~Wang, Y.~Huang, C.~Chen, Z.~Liu, S.~Wang, and Q.~Wang, ``Software testing with large language models: Survey, landscape, and vision,'' {\em IEEE Trans. Softw. Eng.}, vol.~50, p.~911–936, Feb. 2024.

\bibitem{zhao2024arxive}
S.~Zhao, Y.~Huang, J.~Song, Z.~Wang, C.~Wan, and L.~Ma, ``Towards understanding retrieval accuracy and prompt quality in rag systems,'' 2024.

\bibitem{hudson2024softwareeng}
S.~Hudson, S.~Jit, B.~C. Hu, and M.~Chechik, ``A software engineering perspective on testing large language models: Research, practice, tools and benchmarks,'' 2024.

\bibitem{ChangSurvey2024}
Y.~Chang, X.~Wang, J.~Wang, Y.~Wu, L.~Yang, K.~Zhu, H.~Chen, X.~Yi, C.~Wang, Y.~Wang, W.~Ye, Y.~Zhang, Y.~Chang, P.~S. Yu, Q.~Yang, and X.~Xie, ``A survey on evaluation of large language models,'' {\em ACM Trans. Intell. Syst. Technol.}, vol.~15, Mar. 2024.

\bibitem{wang2023measuring}
H.~Wang, J.~Chen, K.~Shu, and L.~Zhang, ``Measuring reliability of large language models through semantic consistency,'' in {\em Proceedings of the 45th International Conference on Software Engineering}, ICSE '23, pp.~982--993, IEEE/ACM, May 2023.

\bibitem{Xuanfan2024}
X.~Ni and P.~Li, ``A systematic evaluation of large language models for natural language generation tasks,'' 2024.

\bibitem{dhuliawala2024}
S.~Dhuliawala, M.~Komeili, J.~Xu, R.~Raileanu, X.~Li, A.~Celikyilmaz, and J.~Weston, ``Chain-of-verification reduces hallucination in large language models,'' in {\em Findings of the Association for Computational Linguistics: ACL 2024} (L.-W. Ku, A.~Martins, and V.~Srikumar, eds.), (Bangkok, Thailand), pp.~3563--3578, Association for Computational Linguistics, Aug. 2024.

\bibitem{Shahul2024}
S.~Es, J.~James, L.~Espinosa~Anke, and S.~Schockaert, ``{RAGA}s: Automated evaluation of retrieval augmented generation,'' in {\em Proceedings of the 18th Conference of the European Chapter of the Association for Computational Linguistics: System Demonstrations} (N.~Aletras and O.~De~Clercq, eds.), (St. Julians, Malta), pp.~150--158, Association for Computational Linguistics, Mar. 2024.

\bibitem{douze2024faiss}
M.~Douze, A.~Guzhva, C.~Deng, J.~Johnson, G.~Szilvasy, P.-E. Mazar{\'e}, M.~Lomeli, L.~Hosseini, and H.~J{\'e}gou, ``The faiss library,'' {\em arXiv preprint arXiv:2401.08281}, 2024.

\end{thebibliography}

\end{document}